# Aging Dynamics in Ferroelectric Deuterated Potassium Dihydrogen Phosphate

Rachel Hecht, Eugene V. Colla, and M. B. Weissman

*Department of Physics, University of Illinois at Urbana-Champaign*

*1110 West Green Street, Urbana, IL 61801-3080*

## ABSTRACT

Anomalously large dielectric aging is found in the high-susceptibility plateau ferroelectric regime of ~95% deuterated potassium dihydrogen phosphate (DKDP). Much less aging is found in non-deuterated KDP. Optical images of the DKDP domain structure show no dramatic change during aging. Small changes in electric field restore the pre-aged susceptibility, but the previous aging almost recovers after returning to the aging field. Susceptibility vs. field can show memory of at least two prior aging fields. Rectifying non-linear susceptibility develops for fields slightly above or below a prior aging field. Aging effects are not fully erased even by brief heating above the Curie point, indicating a role for diffusion of hydrogen to the domain walls, leaving changes in disorder that can survive temporary absence of domain walls. Abrupt random-sign steps in polarization on cooling are accompanied by increases in susceptibility, indicating competition between large-scale domain-wall interaction effects and effects of local interactions with disorder.

**Introduction**

$KH_2PO_4$ (KDP)[1] and its deuterated version, $KD_2PO_4$ (DKDP)[2], are commonly used non-linear dielectrics, each with a ferroelectric (FE) phase. In KDP and DKDP, the FE phases show unusual plateaus in the real dielectric constant $\varepsilon'$ over a temperature range below the Curie temperature, $T_C$, with the large susceptibility arising from a dense array of 180° domain walls.[2-4] The origins of the abrupt loss of susceptibility at low temperature are not fully understood, but they may be driven by a phase transition or sharp crossover, involving interactions among the domain walls[3,5] and probably pinning by interaction with disorder[6,7]. This effect motivates studying dynamical symptoms of such cooperative behavior in the plateau region, such as the aging effects—i.e. reduction in $\varepsilon'$ over time under fixed conditions—that have been found in KDP.[6] Here we report an unusually strong aging effect in a typical, incompletely deuterated[2,8] commercial DKDP sample. In contrast, KDP shows much smaller aging. Since both materials have similar plateau regions, the origins of the large aging in DKDP shed only indirect light on the physics of the plateau. The origin of the large aging is also interesting in itself.

There are, broadly speaking, two main types of aging in disordered domain systems.[9] One consists simply of domain growth, which reduces the domain-wall area, and thus reduces the large component of $\varepsilon'$ that comes from the domain walls. The other consists of gradual settling of domain walls into low free-energy configurations, stuck to underlying disorder. In typical cases, this settling consists of changes in domain wall configuration to fit quenched disorder[6,9], but it can also result from changes in the disorder to fit a particular domain wall configuration. The latter case is often found in magnetic after-effects (e.g.[10], and some ferroelectric systems, e.g.[11]), driven by the diffusion of mobile defects to domain walls. Since for KDP and DKDP interactions among the domain walls are suspected to lead to a sort of glassy freezing[3,5], in addition to traditional domain aging mechanisms there could also be more complicated aging typical of glasses and spinglasses, but with domains as the ingredients.[9]

We use a variety of probes, ranging from optical imaging to behavior of $\varepsilon'$ and non-linear susceptibility under various field-temperature (E, T) histories, to probe the origins of what turns out to be anomalously large aging in DKDP. The main types of aging under consideration include:

1) Simple domain growth, i.e. reducing the net area of the domain walls from which most of the response arises.
2) Settling of the system of interacting domain-walls into a low free-energy state, with reduced response in the experimental frequency range.
3) Settling of the domain walls and lattice disorder into a well-pinned low free-energy state, with reduced response in the experimental frequency range. This can occur by either
    a. detailed rearrangement of domain-wall configuration to fit fixed lattice disorder patterns,
    or
    b. diffusion of disorder to fit domain walls.

We first confirm previous reports[2,3,5] that DKDP and KDP have very similar dynamical freezing at the low-T edge of the plateau. Thus the large aging found in DKDP is not directly due to the shared physics behind that effect, whatever it may turn out to be. Likewise, we find major non-linear response in DKDP, similar to that previously reported in KDP.[7] We then present results showing that the aging effect in DKDP is not due to domain growth, and that the settling of interacting domain walls into the frozen set found below the plateau competes with rather than enhances the aging, leaving domain-wall pinning effects as the mechanism.

The distribution of H and D in partially deuterated KDP should be an important source of disorder, because $T_C$ increases as a function of D concentration.[2,4,8] In equilibrium, D should preferentially sit in the regions with good ferroelectric order, leaving H preferentially at the domain walls. In KDP, the deuterium concentration should be about the natural abundance, 0.016%, so that H-D disorder would be much less important than in our DKDP samples. Thus interaction between the domain walls and the large disorder due to incomplete deuteration is the obvious suspect for the cause of the large aging we shall describe in DKDP.

This aging effect can occur either by domain wall configuration adjustments or by relocation of H to domain walls. For partially deuterated KDP, it is known that the H-D distribution diffuses, with typical diffusion coefficients at room temperature on the order of $10^{-15}$ cm$^2$/s, varying as a function of D concentration.[12] Although we have found no studies of the temperature dependence of this rate, the diffusion was interpreted as being via tunneling, so extreme temperature dependences would not be expected.[12] Since we shall show that the aging can partially survive the temporary melting of the domain walls, we shall conclude that diffusion of H to domain walls is a key mechanism.

**Materials and Methods**

DKDP crystals were obtained from United Crystals and KPD crystals were grown from water solution at the University of Illinois at Urbana-Champaign. For optics, the samples were polished using lapping discs with grit size down to 0.3 µm. Optical images, including digital movies, of the domain structure were obtained via polarized light microscopy using a Leica DM2700 microscope. Typical dimensions of the samples used for dielectric measurements were ~ 1mm thickness and ~ 6.5mm$^2$ area. Samples were made into capacitors by depositing a thin layer of Cr (~10nm) followed by about 100 nm of Ag via thermal evaporation.

Most of our dielectric response measurements were made with applied ac voltage at 100Hz, using a standard lock-in amplifier with a circuit that allowed dc voltage to be applied along with the ac voltage. Measurement of dielectric response for KDP is complicated by large non-linear effects, attributed to weakly pinned domain walls[7], found at relatively low fields. Fig. 1 illustrates the voltage-dependence of the response for DKDP, showing effects very similar to those reported for KDP.[7] Most of our measurements were made using 1V rms or, for the second-order non-linear response, 2V rms—large enough to be well into the non-linear regime in which several frequency-dependent features show up that are not apparent in the linear regime, and for which aging effects are particularly evident.[7] Broad-band dielectric response data were taken using QuadTech 7600 Plus LCR meter using 1 V rms ac bias, and thus also well into the non-linear regime. Although the response measured is therefore not, strictly speaking, the linear dielectric coefficient, we will still refer to the apparent coefficient determined from the current/voltage ratio as ε' for brevity.

**Results**

All samples showed the usual FE transition with a plateau regime for ε', as shown in Fig. 2. Analysis of the temperature-dependence of the characteristic peak frequencies in ε"(f,T) shown in Fig. 3 shows nearly-ordinary Arrhenius behavior near the high-T end of the plateau and sharply non-Arrhenius behavior near the low-T end of the plateau, for which no plausible Arrhenius attempt rate could be used, similar to previous reports[3,5]. (Previous work shows that the characteristic frequencies of these peaks in KDP are only very weakly dependent on ac field amplitude.[7]) This non-Arrhenius freezing, following a Vogel-Fulcher form, confirms that a phase transition—or at least a strongly cooperative crossover effect—occurs at the low-T side of the plateau.

Fig. 4 shows ε' aging vs. time for DKDP at E=0 at several temperatures. (The initial oscillations are due to the temperature controller.) Very substantial aging is found throughout the plateau region. Fig. 5 contrasts the large aging in DKDP within the plateau region with the smaller aging in KDP. The fractional rate at which ε' decreases depends not only on $T_{aging}$ but also on whether the aging run is preceded by annealing at 400K or at 300K, as indicated in Fig. 5a, showing the fractional decrease in the first 5 hours of aging.

It is evident that the large aging in DKDP cannot be a direct effect of the physics shared with KDP, which shows much smaller aging. Thus our further experiments concentrated on the large, easily measured, previously unreported aging in DKDP. The substantial difference of the aging after between-run anneals to 300K and 400K (well above $T_C$) strongly suggests that some ingredient other than domain walls and fixed disorder will be required, since fixed disorder doesn't anneal and all domain-wall effects should completely reset near $T_C$.

Polarized light microscopy was used to directly visualize the aging domains. Figure 6 shows the domain structure during the aging at T=200K. Over most of the sample the approximate domain size is unchanged, and there are only small motions of the domain walls. In one anomalous region (lower part of the picture) the domain structure changed substantially. Since the approximate aging behavior was similar in each sample and involves huge fractional changes in $\varepsilon'$, we believe that the anomalous regions are not needed to understand its crudest qualitative features. We shall, however, see relatively large variability between samples and on repeated thermal cycles. Large defects in the domain pattern are likely to be involved in those effects.

Fig. 7 shows the response of the aging after changes in E, most notably rejuvenation after sufficient change in E. Despite the rejuvenation after changing E, most of the memory of aging is restored fairly quickly after resetting to the initial value of E. This behavior seems hard to reconcile with a picture of domain walls adjusting their complicated, detailed configuration to quenched disorder, since a field change sufficient to erase the previous aging should move most domain walls too far from their previous positions to retain such memory. On the other hand, a simple potential well formed by accumulated H would remain in place at the original domain wall positions, ready to affect the response once the domain walls returned to their original position when the initial E is restored.

Fig. 8 shows ε' and ε" as a function of E on fairly rapid E-sweeps taken after aging at 185 V/cm at 202K. These show a clear hole in ε' and a peak in ε", with characteristic half-widths of about 19 V/cm. As discussed below, that width corresponds approximately to the field change needed to move domain walls about one domain-wall width. The persistence of memory even after E excursions well outside the range of the memory hole again appears inconsistent with a picture exclusively of detailed adjustment of domain walls to quenched disorder; but it is consistent with adjustment of the disorder to fit the domain wall positions found at the aging E. Some hysteresis appears in the position of the aging hole, depending on the sign of dE/dt. An effect of this sort is expected in an H-diffusion picture, since the recent history of the domain wall positions not only creates new inhomogeneities of H but can also drag old ones around. (We have not attempted to calculate the magnitude or even the sign of that effect.) One peculiar feature is that the peak in ε" is absent immediately after aging and only shows up on the subsequent sweeps through $E_{aging}$, unlike the hole in ε'.

Fig. 9 shows ε'(E) data similar to Fig. 8, but taken after aging 10 hours at 185 V/cm and then 10 hours at 370 V/cm. Two clear holes appear, so the system can remember at least two different E fields. This result is compatible with aging creating sheets of enhanced H concentration, but difficult to explain in a picture of detailed domain wall adjustment to fixed disorder.

The dependence of ε' on E implies a non-zero second derivative of polarization P with respect to E, which would lead to second-harmonic generation whose sign would depend on the sign of the change in E after aging. Fig. 10 shows this effect. There is also an offset in the second harmonic generation coefficient depending on the sign of the sweep of E(t). This effect is not obtainable from simply taking the derivative of the ε' shown in the Fig. 8 with respect to E. It would nonetheless be expected, because in addition to the deliberate long-time aging, there will be ongoing short-time aging, leaving each domain wall in an asymmetrical potential as E is swept over time, with the sign of the asymmetry depending on the direction of the sweep.

Fig. 11 shows aging similar to that in Fig. 6, but with a brief interruption after 10 hours to heat to 211.5K—just above $T_C$. Remarkably, this heating leaves the aging effect largely in place, indicating that something other than the domain walls themselves (which are lost at $T_C$) must carry the memory. This is a particularly strong piece of evidence pointing to the formation of H sheets stabilizing domain walls, since such sheets would only gradually diffuse away above $T_C$. The partial loss of memory probably indicates that when a new set of domains form, they only partially follow the pattern of the previous set, despite the pre-existing H-sheets.

Fig. 12a shows the behavior of $\epsilon'(T)$ upon further cooling and then heating after a 5-hour aging pause at 205K during the initial cooling. Under these conditions, there is substantial thermal rejuvenation, i.e. loss of the effects of aging, on cooling a few K below the aging temperature, similar to KDP measured at low ac fields[6]. This loss is not recovered upon reheating to the aging temperature, which contrasts the behavior of spinglasses[9] and some cubic relaxor ferroelectrics[13]. The latter show hole-like memory effects of past aging temperatures, similar to the memories of past fields shown in figures 8-10. Figure 12b, in contrast, shows a similar experiment on another sample with a 20-hour aging, which gives little or no thermal rejuvenation. Here, unlike previous reports on the smaller aging of KDP[6], most of the effect of aging persists throughout the plateau regime and remains upon re-heating to the aging temperature and even above, as seen in the comparison of the after-aging curve with the no-aging reference curves. Although that thermal behavior would be characteristic of simple domain growth, as we have seen, the behavior under electric field changes does not fit any such model.

The data shown so far generally point to a relatively simple electric after-effect picture, in which domain walls become decorated with diffusing H, forming potential wells. One subtlety, however, requires other physics, potentially relevant to understanding the plateau. The rejuvenation on cooling below the aging temperature (Fig. 12) would not arise for domain walls remaining in fixed positions. Small changes in their configuration as a function of T would lead to such an effect, since many domain walls would no longer sit at the positions at which they were especially pinned by the aged disorder. Such changes appear to be a non-equilibrium effect, since the rejuvenation effect is not retraced on re-heating, unlike similar effects in spinglasses[9] and some cubic relaxor ferroelectrics. [13]

The rejuvenation after large-scale domain-wall reconfiguration can be seen particularly clearly when that rearrangement occurs in abrupt quakes. Direct measurement of the current generated by the capacitor, $I_P(t)$, almost always shows a remarkable noise effect during cooling, consisting of large random-sign current spikes. An example run showing such spikes is shown in Fig. 13. The change in polarization of the largest spikes exceeds 1% of the saturation polarization of the sample. Any such current spikes would have to be accompanied by domain wall reconfiguration. If the aging consists of detailed local mutual adjustment of domain walls and disorder to maximize pinning, those domain walls involved in any current spike should rejuvenate. Detailed examination of the change in ε' during $I_P(t)$ spikes, as shown in Fig. 14, shows just such rejuvenation regardless of the sign of the spikes. Fig. 15 shows a warming run in the temperature range for which spikes were densest during cooling. Very few spikes in $I_P(t)$ and no large steps in ε'(t) are found. This behavior was consistent on all warming runs.

**Discussion**

The large aging in DKDP can be accounted for by a simple picture of H diffusion to the domain walls, creating free-energy wells. The key quantitative feature, the characteristic field scale for rejuvenation, is consistent with pictures of aging driven by local adjustments of the domain walls and the disorder. The polarization change for a 230 V/cm field corresponds to ~ 0.2% of the saturation polarization, corresponding to domain wall motion of ~0.1% of the domain-wall spacing. Based on the optical images, that would be ~5nm displacement, which is comparable to the domain-wall thickness measured by X-ray methods.[14] The ability of the system to retain memory of prior aging fields even after much larger field excursions is compatible with the main mechanism being the formation of H-rich sheets, and seems incompatible with a model of detailed domain wall adjustment to fixed disorder. The persistence of memory even after heating to above $T_C$ is dramatic evidence supporting the same conclusion, since it requires formation of some pattern that persists even in the absence of ferroelectric domains.

One detailed oddity should be mentioned: the presence of a peak in $\varepsilon''(E)$ near the aging E on subsequent E-sweeps but not immediately after the aging before a change in E. The change in $\varepsilon'$ is due to both changes in the kinetics and direct growth of the potential well, possibly allowing and accidental cancellation of the effects on $\varepsilon''$ under these conditions. It might be possible to sort out these effects by comparing memory effects on $\varepsilon'$ and $\varepsilon''$ as well as checking for the effects of non-linearity, but that is beyond the scope of this paper.

The aging, rejuvenation, and noise effects also shed light on the physics of the dielectric freezing at the low-T edge of the plateau. Below ~165K aging has very little effect on $\varepsilon'$, although above ~160K there is still a large $\varepsilon'$ associated with the plateau. Since after sufficient aging time the aging effect mostly retraces upon subsequent warming, the inhomogeneities in the H distribution remain below 165K but have little effect on $\varepsilon'$. It seems that at low T the domain walls are held in place by another relatively strong potential, such as the domain wall interactions. The irreversible rejuvenation found on cooling after 5 hours of aging may indicate that this competing term in the Hamiltonian is strong enough to move many of the walls out of their aged positions. The different rejuvenation behavior found under slightly different conditions apparently reflects competition between the potential energy of the wells formed by aging and the other term in the potential energy. If the former is not strong enough, essentially all the walls get moved around on cooling to the low-T end of the plateau, giving rejuvenation. Since the relative size of the competing effect grows at low T, that indicates that it comes from the same cooperative physics that gives rise to the highly non-Arrhenius freezing, most likely arising from domain-wall interactions.

The noise effects are consistent with the same picture of competing short-range disorder and long-range interaction effects. Domain quakes occur almost exclusively upon cooling. As the strength of the cooperative domain-wall interaction term grows, walls jump from local minima to new global minima, restarting the process of developing local aging minima. The response of $\varepsilon'$ to a quake thus represents rejuvenation of part of the sample. Upon warming, in contrast, the strength of the cooperative interaction terms weakens, so domain walls that are in some metastable global minimum remain in the same global minimum, developing local minima without interruption.

**Acknowledgments**. R. Hecht was supported by NSF REU grant 13-59126. We thank A. J. M. Grinberg for able technical help with the 2f measurements.

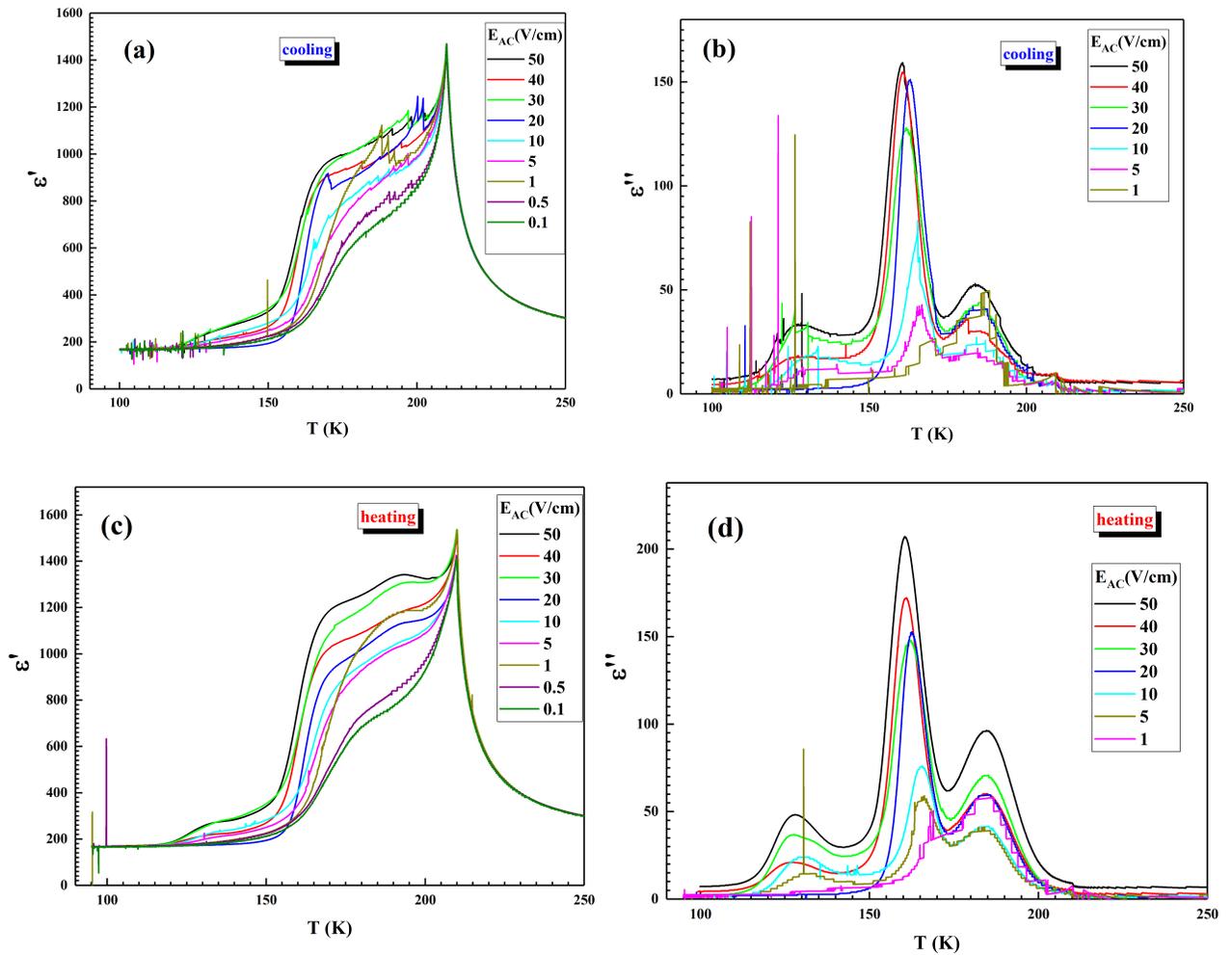

Fig. 1. DKDP dielectric susceptibility measured with different ac drive magnitudes. (a) (ε') and (b) (ε") are measured during cooling at 4K/min and (c) and (d) – during heating at 4K/min. The ε" panels omit data for 0.5 and 0.1 V/cm, which look just like the 1V/cm data except with more instrumental noise. Sharp spikes appearing often in the cooling data and rarely in the heating data do not represent actual changes in ε', but are caused by spikes in the polarization current. Triangle-like steps, however, do represent changes in ε'.

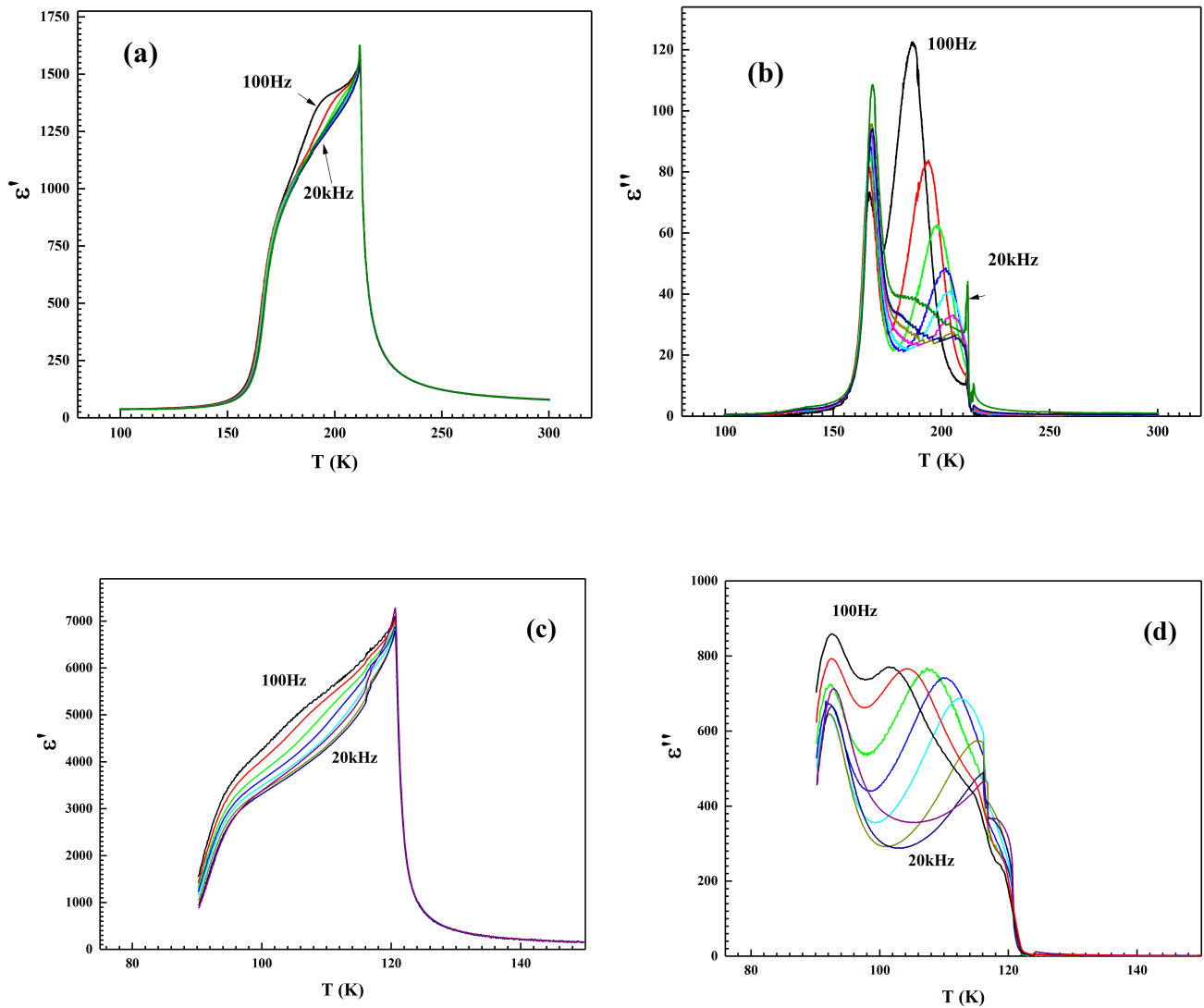

Fig. 2. Temperature dependencies of the real and imaginary parts of dielectric susceptibility of DKDP (a,b) and KDP (c,d) measured with 1 V rms ac drive at frequencies from 100 Hz to 20 KHz. Measurements were made on cooling at 4K/min.

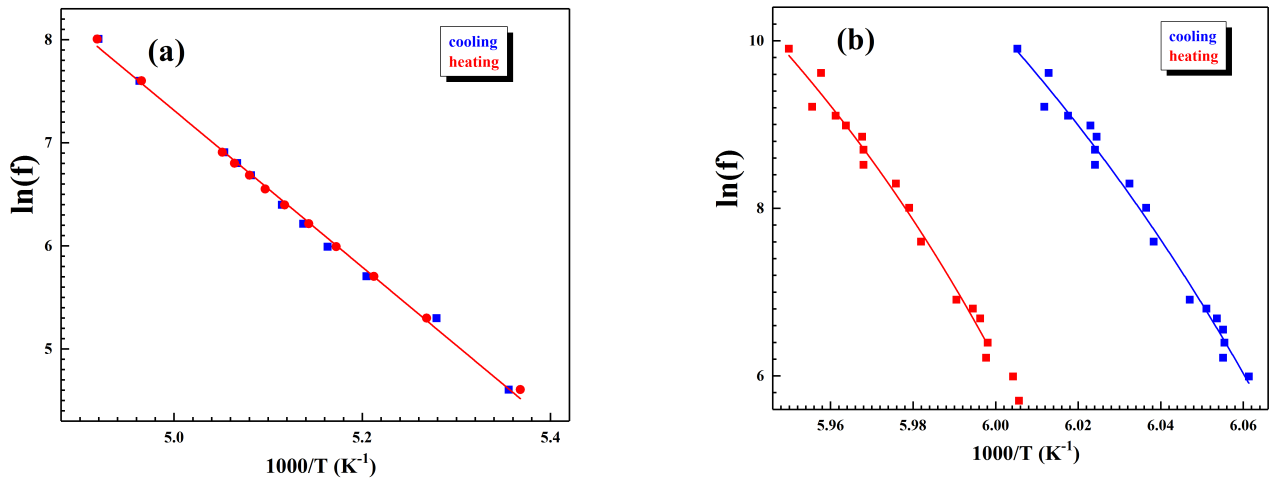

Fig. 3 shows nearly-Arrhenius frequency dependence of peak temperatures near $T_C$, and much steeper dependence of frequency on T near the lower end of the plateau. These measurements were made using fixed frequency and swept T. The results in this range are close to those obtained using fixed T and swept f. The peak near $T_C$ shows an attempt rate of $10^{20}$ Hz, mildly unphysical, indicating gradual growth of barriers on cooling. The peak near 165K shows distinct curvature from the Arrhenius form, with an apparent Vogel-Fulcher temperature near 160K, and reasonable attempt rates (very sensitive to the Vogel-Fulcher parameter) of very roughly $10^{10}$ Hz.. An Arrhenius fit would require an extremely unphysical attempt rate above $10^{160}$ Hz.

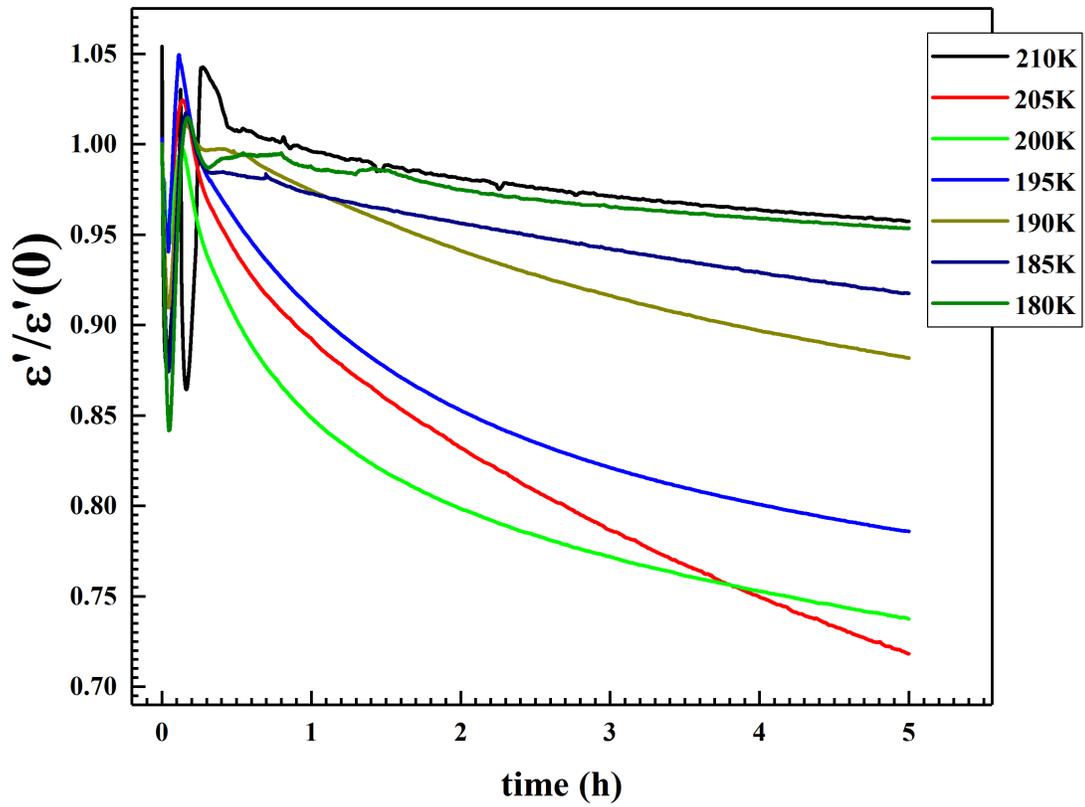

Fig. 4 shows aging in ε'(t) for a few temperatures for DKDP. Oscillations in the first half hour are due to the temperature controller.

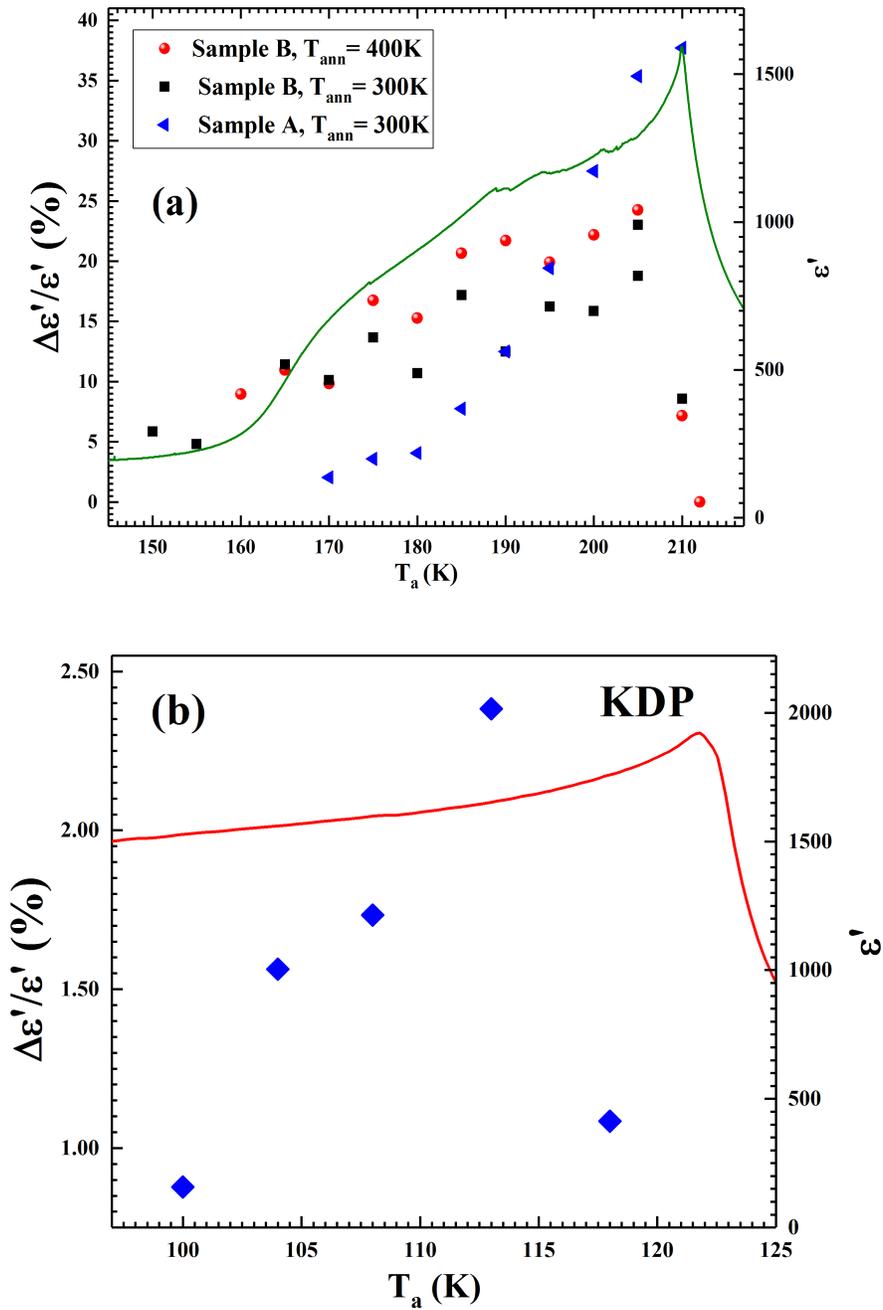

Fig. 5. The points show the fractional reduction in ε' in the 5 hours after reaching the temperature shown on cooling for two DKDP samples (a), and a KDP (b) sample, with the lines serving as a reminder of ε'(t). No aging is found above $T_C$. The aging fraction depends on sample and on annealing temperature between runs.

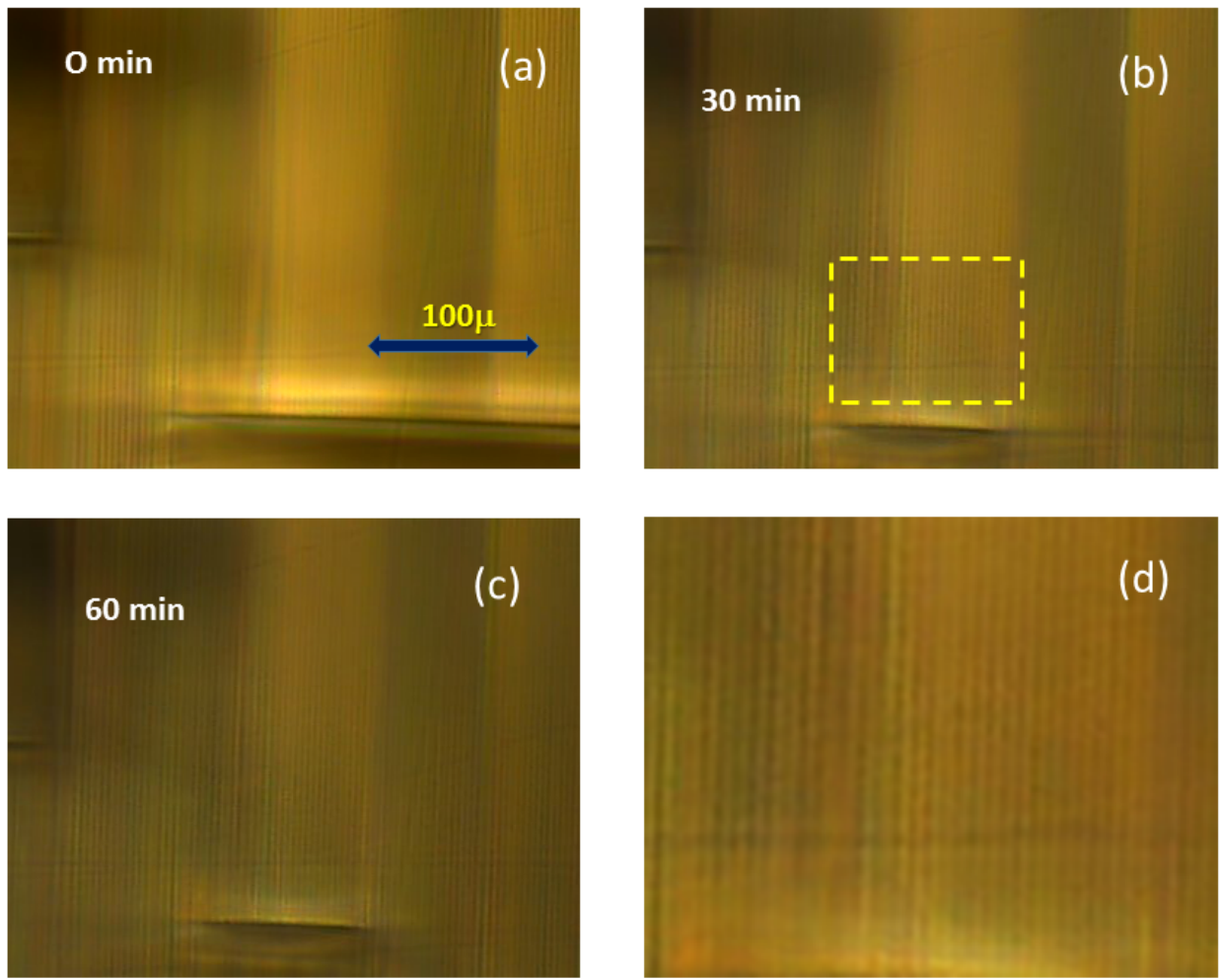

Fig. 6. Evolution of the domain pattern during the aging in a DKDP sample at 200K, viewed by polarized light microscopy. Panel (d) shows a blow-up of the marked region from panel (b), in which the near-vertical stripes representing the main domain structure are clearly visible.

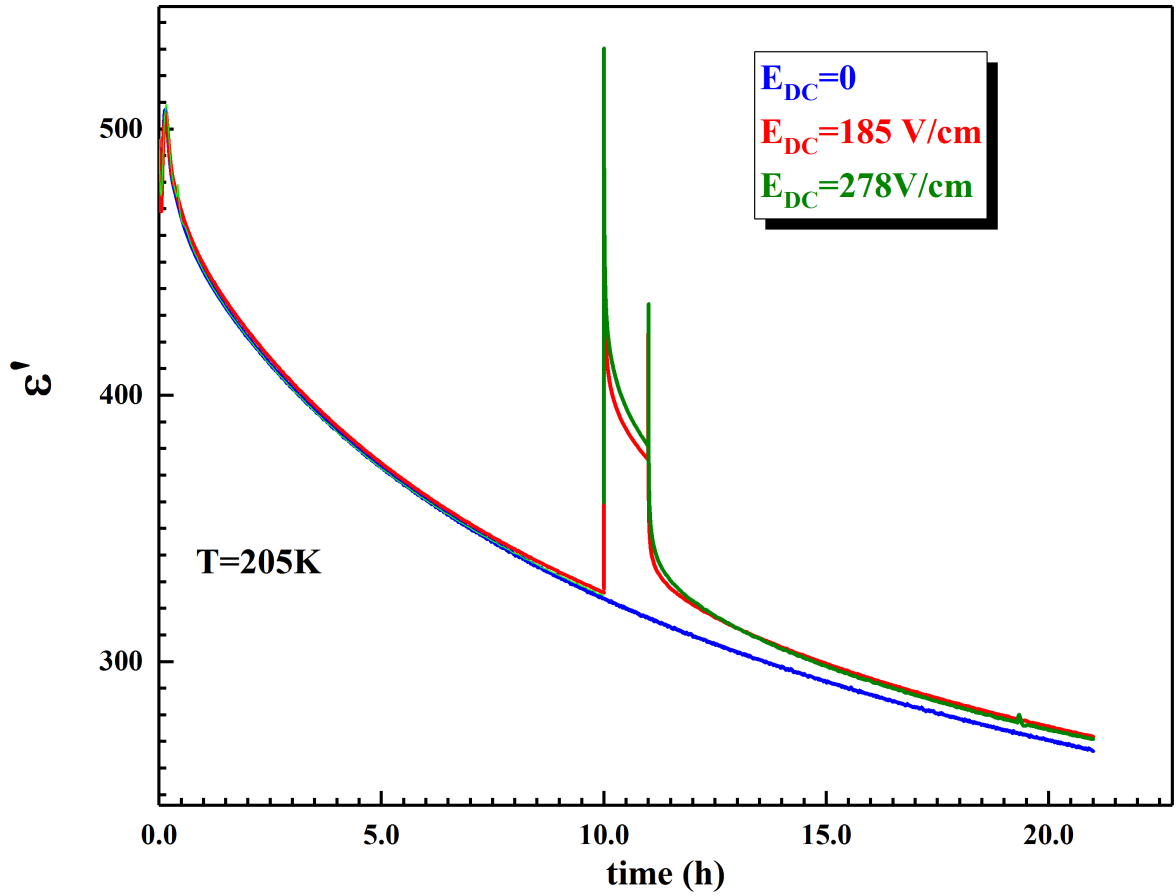

Fig. 7. Aging of ε' for 21 hours at T=205K (a), with E=0, and with 1hr interruptions of with the dc fields of 185 V/cm and 278 V/cm.

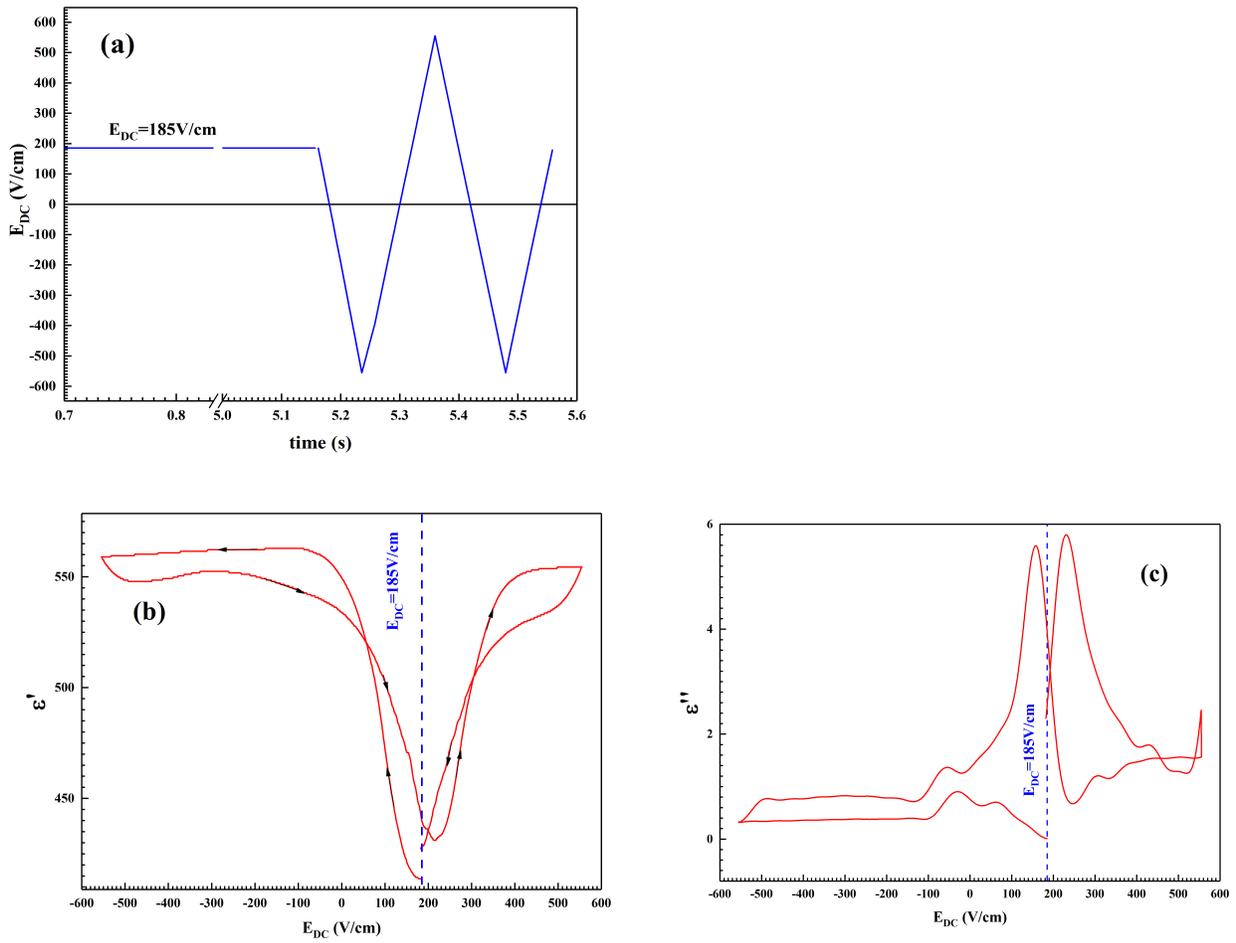

Fig. 8. (a) Shows the DC bias protocol for the "hole-burning" experiment, with T=202K. (b) Shows the ε'(E) measured during the field-sweep parts of the protocol after aging at $E_{DC}$=185V/cm, and (c) shows ε", smoothed to reduce noise.

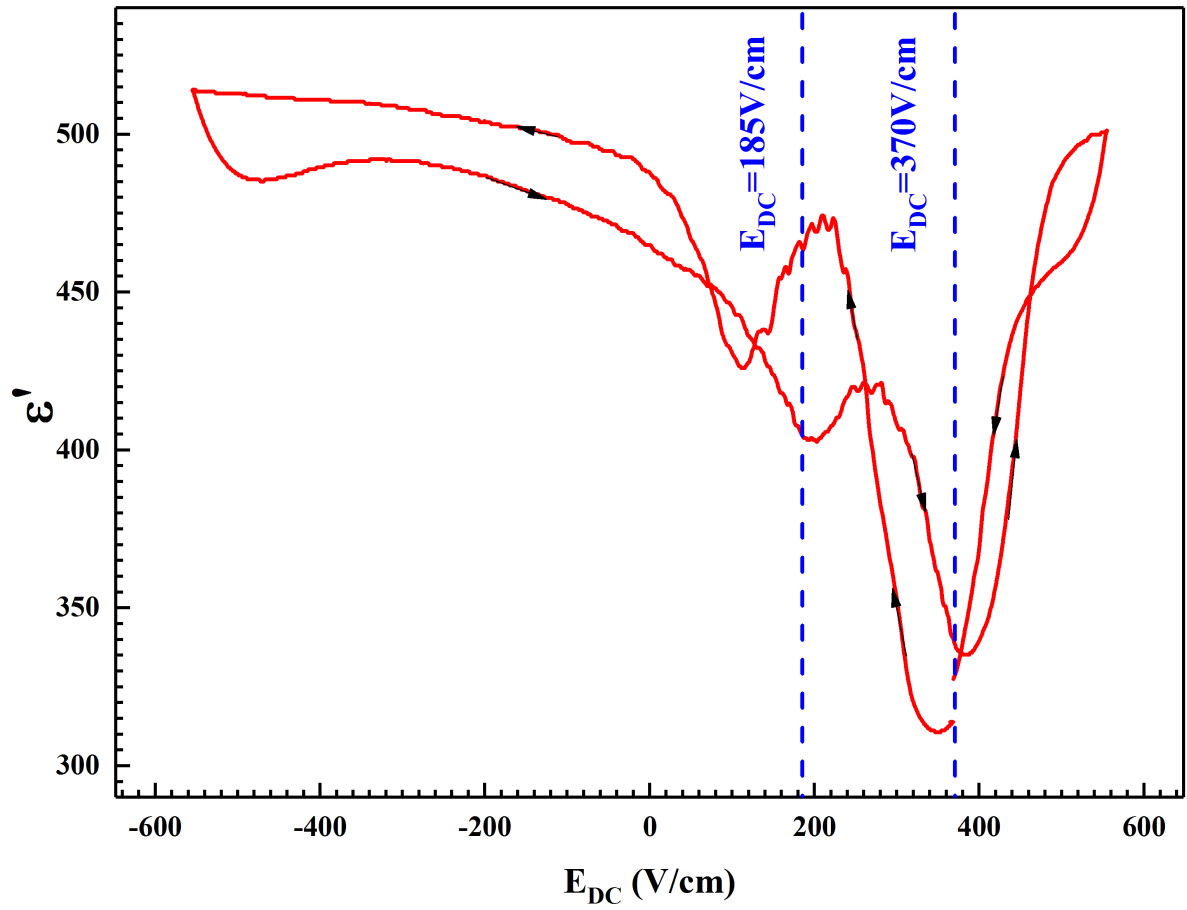

Fig. 9. Aging in ε' at 202K with $E_{DC}$=185V/cm for 10 hours followed by $E_{DC}$=370V/m for 10 hours, then measured by scanning $E_{DC}$ as shown in Fig. 8.

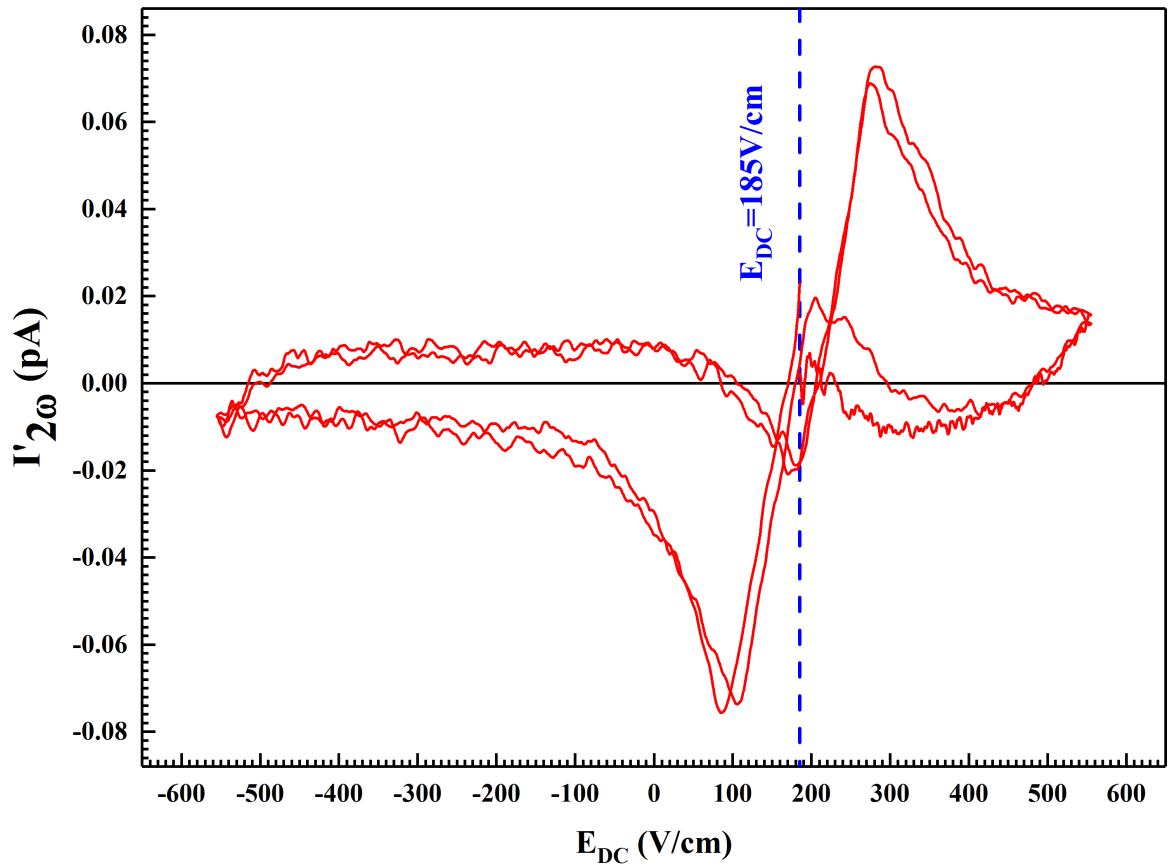

Fig. 10. Second harmonic current component in DC bias hole-burning experiment of Fig. 8. The two similar traces were taken consecutively, as shown in the protocol of Fig. 8a. Positive sign here means positive current maxima at the maxima of $(dV/dt)^2$ of the fundamental ac voltage drive, as would be produced by an instantaneous quadratic response. The out-of-phase second harmonic current was much smaller.

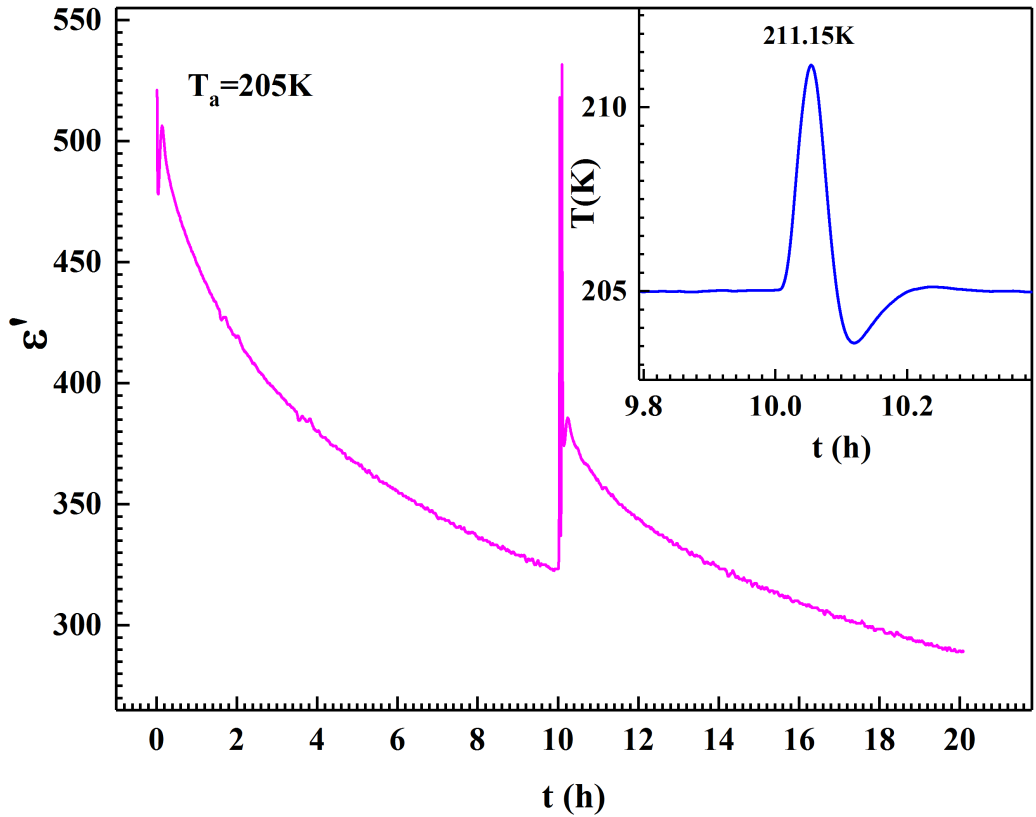

Fig. 11. Aging at 205K interrupted by a temperature excursion to 211.5K. As in these results, a significant portion of the aging from the first 10 hours survived heating above $T_C$ in similar trials using maximum temperatures of up to 216.5K. In each such trial, $\varepsilon'$ itself showed that the sample was heated above the peak at $T_C$, serving as a direct local thermometer.

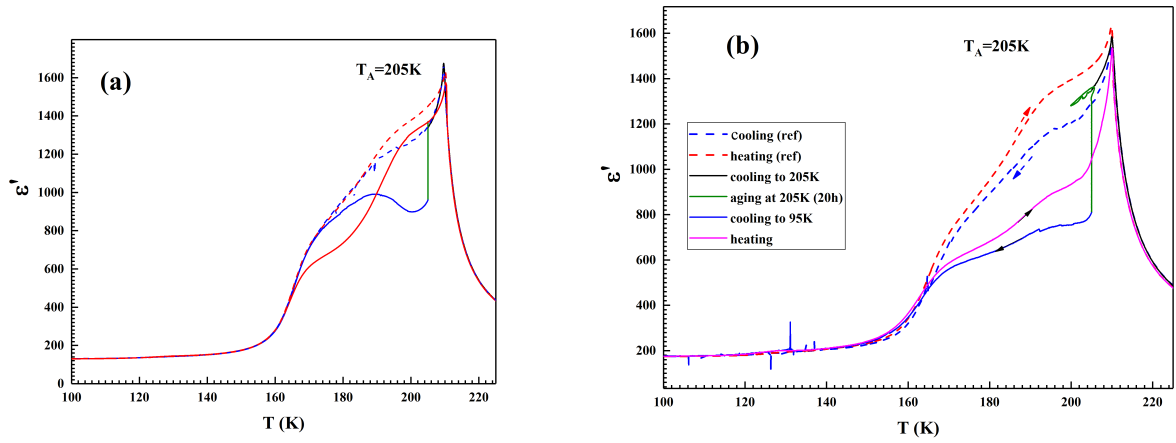

Fig. 12. (a) shows ε' on a temperature-sweep of sample A with a 5 hour pause for aging at 205K during cooling, compared with a reference without a pause. The sweep rate is 4K/min. The dashed lines represent the reference susceptibility obtained while cooling (blue) and heating (red) with no aging. (b) shows ε' on a T-sweep of sample B with a 20 hour pause for aging at 205K during cooling, compared with a reference without a pause. The sweep rate is 4K/min. The smaller upward steps on cooling are real, not instrumental artifacts. Random-sign spikes at low temperature are from spikes in $I_P$, not actual changes in ε'.

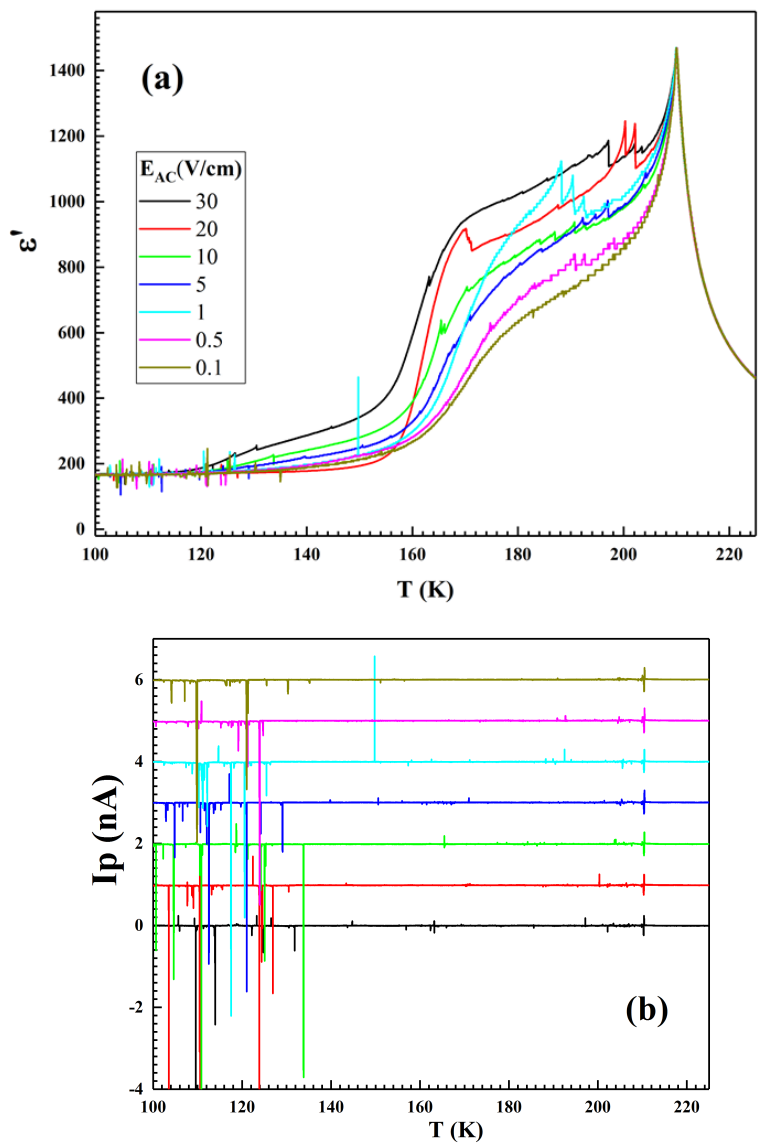

Fig. 13. (a) Steps in ε' and (b) spikes in $I_P$ upon cooling at several different ac bias voltages. (Apparent random-sign spikes in ε' are artifacts produced by the spikes in $I_P$.)

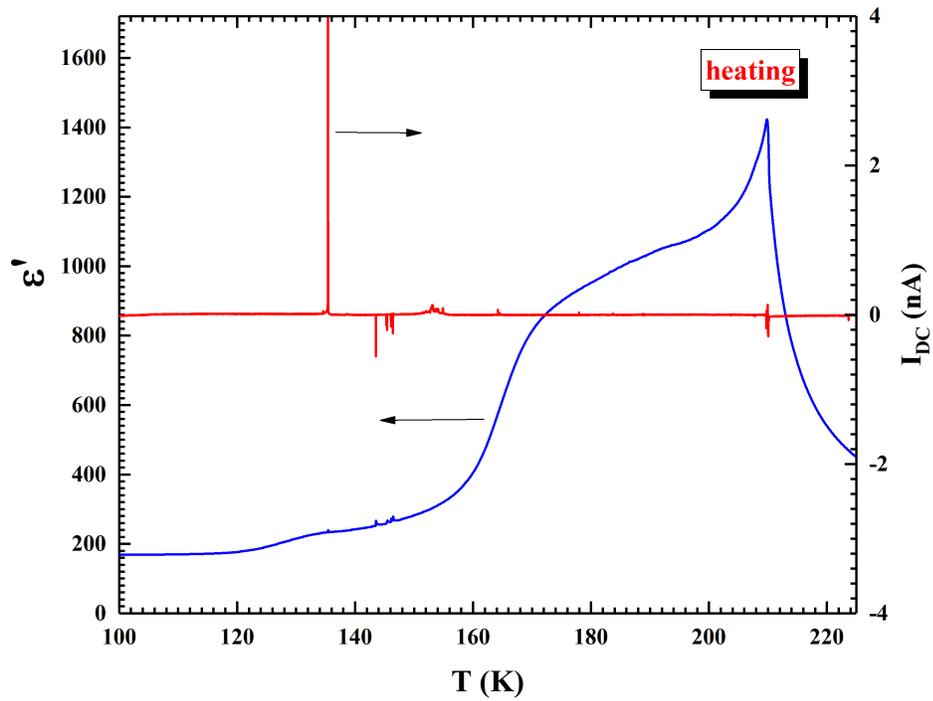

Fig. 14. A record of $I_P$, similar to that of Fig. 14, but taken during heating. $\varepsilon'$ is shown for the same sweep with $E_{AC}$ 10V/cm.

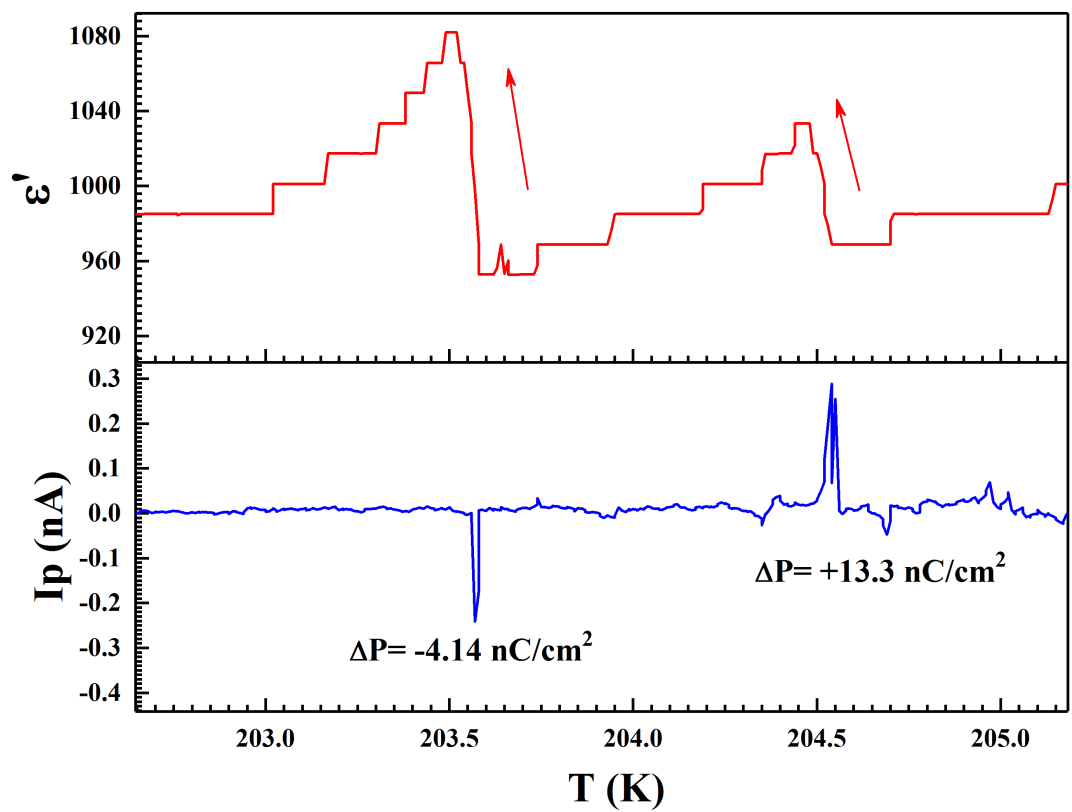

Fig. 15. Random-sign $I_P$ spikes found on cooling are accompanied by upward steps in $\varepsilon'$.